\documentclass[12pt,preprint]{aastex}

\usepackage{epsfig}

\begin{document}

\newcommand{\vdag}{(v)^\dagger}
\newcommand{\myemail}{fukumura@physics.montana.edu}
\shorttitle{Iron Lines from Spiral Waves} \shortauthors{Fukumura
et al.}

\title{Iron $K \alpha$ Fluorescent Line Profiles from Spiral Accretion Flows in AGNs}

\author{Keigo
Fukumura\footnote{http://www.physics.montana.edu/students/keigo/homepage/html/research.html}
and  Sachiko Tsuruta} \affil{Department of Physics, Montana State
University,
       Bozeman, MT 59717-3840;
       \myemail, tsuruta@physics.montana.edu}

\begin{abstract}
We present 6.4 keV iron $K \alpha$ fluorescent line profiles
predicted for a relativistic black hole accretion disk in the
presence of a spiral motion in Kerr geometry, the work extended
from an earlier literature motivated by recent magnetohydrodynamic
(MHD) simulations. The velocity field of the spiral motion,
superposed on the background Keplerian flow, results in a
complicated redshift distribution in the accretion disk. An X-ray
source attributed to a localized flaring region on the black hole
symmetry axis illuminates the iron in the disk. The emissivity
form becomes very steep because of the light bending effect from
the primary X-ray source to the disk. The predicted line profile
is calculated for various spiral waves, and we found, regardless
of the source height, that: (i) a multiple-peak along with a
classical double-peak structure generally appears, (ii) such a
multiple-peak can be categorized into two types, sharp sub-peaks
and periodic spiky peaks, (iii) a tightly-packed spiral wave tends
to produce more spiky multiple peaks, whereas (iv) a spiral wave
with a larger amplitude seems to generate more sharp sub-peaks,
(v) the effect seems to be less significant when the spiral wave
is centrally concentrated, (vi) the line shape may show a drastic
change (forming a double-peak, triple-peak or multiple-peak
feature) as the spiral wave rotates with the disk. Our results
emphasize that around a rapidly-rotating black hole an extremely
redshifted iron line profile with a noticeable spike-like feature
can be realized in the presence of the spiral wave. Future X-ray
observations, from {\it Astro-E2} for example, will have
sufficient spectral resolution for testing our spiral wave model
which exhibits unique spike-like features.
\end{abstract}


\section{Introduction}
It is generally known that Seyfert 1s spectra exhibit reprocessed
X-ray emission, such as broad Fe K$\alpha$ fluorescence at 6.4 keV
and Compton reflection (high energy hump) around $\sim 30$ keV,
from an illuminated accretion disk around a supermassive black
hole. If the origin of such a reflection component can be
attributed to a surrounding accretion disk {\it (disk-line
model)}, more precise spectral observations might allow us to
better understand the physics in the vicinity very close to the
central black hole. In standard disk-line models
\citep{Fab89,Laor91,Kojima91}, accreting matter is normally
considered to be in Keplerian orbit around a central black hole.
The disk is then illuminated by an external X-ray source located
somewhere near the innermost disk region and produces a broad,
skewed fluorescent emission line at a particular photon energy
(6.4 keV for neutral-like iron, 6.7 keV for He-like iron, and 6.9
keV for H-like iron, for instance) depending on the ionization
state. See \cite{Fab02} and \cite{Reynolds03} for comprehensive
review.

Past X-ray observations with the Japan/USA X-ray satellite {\it
ASCA} (the best resolving power of $\sim 120$ eV) revealed such a
relativistically broadened iron line from various Seyfert 1
galaxies, particularly from a very bright source MCG-6-30-15
\citep{Tanaka95,Fab95}. The observational data seems to be
well-fitted with the standard disk-line model in terms of the
characteristic double-peak feature, which is thought to be due to
the longitudinal Doppler motion of the emitting matter in the
disk, and the asymmetry in the profile is also believed to
originate from the special relativistic, directional beaming
effect. Furthermore, the observed emission line occasionally
contains a very extended red tail that is interpreted to be a
consequence of the general relativistic effect (i.e.,
gravitational redshift).

Apart from the standard disk-line models in which an axially
symmetric flow with Keplerian circular motion is assumed under no
perturbations, some authors have argued other accretion theories
from different aspects. According to the past literatures, there
seem to be three distinct issues of studies involved with such
modifications: those (i) focusing on modified disk geometry
instead of a simplified thin disk in the equatorial plane, (ii)
more complex X-ray source geometry, and (iii) perturbed accretion
flow (i.e., non-Keplerian flows). These issues (i)-(iii) seem to
be independent of each other but they all modify the standard line
profiles in terms of the redshift of photons, the line shape,
and/or the line variabilities. As far as issue (i) is concerned,
the actual disk geometry may not be so simple as
geometrically-thin in the equatorial plane. For instance,
\cite{HB00} has discussed warping of the disk where a part of the
disk surface is geometrically skewed and tilted from the
equatorial plane. On the other hand, a thin disk might become
hotter as the gas approaches the hole, forming a
geometrically-thick disk (or an accretion torus). Such a thick
disk model (accretion torus) has been studied by a number of
authors, e.g., \cite{SLE76}, \cite{TT89}, and \cite{KF92}, while
dense clouds embedded in a thick disk has been discussed by
\cite{Guilbert88}, \cite{Sivron93} and \cite{HB01}. The second
standpoint (ii) includes a stationary/dynamical, off-axis X-ray
source (i.e., a non-axisymmetric point source) instead of a
typically postulated axisymmetric X-ray source. That is, the
primary source (for example, an orbiting flare in a corona or high
energy particles due to shock accelerations) is either located
off-axis or moving around with some bulk motion relative to the
disk \citep{RB97,Rus00,DL01,LY01}. A flaring activity induced in
the disk itself might irradiate the Keplerian matter to produce
iron fluorescence \citep[]{Nayakshin01}. They found a very narrow
line owing to the small height of the flare.

With regard to the third issue (iii), Keplerian motion may not be
a very practical assumption since a thin disk is known to be
subject to various kinds of instabilities to perturbations. For
instance, disk oscillations can be triggered in the form of the
(p,g,c)-mode in the innermost disk region \citep{kato01a,kato01b},
in which case the local emissivity is no longer axisymmetric
\citep[][]{YF93,sanbuichi94,karas01}. Also, magnetohydrodynamic
(MHD) instabilities such as the magneto-rotational instability
(MRI) can induce spiral density waves in the presence of the
magnetic field. This has been phenomenologically confirmed by
two-dimensional numerical studies of a magnetized disk in
Newtonian gravity where a multiple-armed spiral wave evolves into
a single-armed spiral over some time \citep[hereafter CT01]{CT01}.
On the other hand, a recent three-dimensional MHD simulation of
the black hole accretion by \citet[hereafter MM03]{Machida03}
suggests that an X-ray flare via magnetic reconnections could take
place in the plunging region (i.e., the region inside the
marginally stable orbit) as a result of the non-axisymmetric
(one-armed) spiral density structure, which is initially caused by
MRI due to a differential rotation of frozen-in plasma.
\citet[hereafter H01]{Hawley01} has independently predicted a
similar formation of tightly wrapped spiral waves using the
Pseudo-Newtonian potential. It is also demonstrated by
\cite{Armitage02} that the intensity of the magnetic fields in the
accretion disk tends to be amplified by shearing, which could be a
potential source of the observed X-rays. Additionally, a
gravitational perturbation (tidal force) by a nearby star or a
massive gas can trigger such a spiral wave which eventually
steepens into shocks \citep{cha93a}. The line profiles under such
spiral patterns have been found by several authors
\citep{cha93b,cha94} in the Newtonian geometry. In all these
cases, the accreting flow is no longer Keplerian because of a
substantial radial velocity component. Therefore, the velocity
field should be somewhat deviated from a circular one. For
instance, \citet[hereafter HB02]{HB02} considered the possible
effects of such a spiral motion on the effective line profile in
an approximate Schwarzschild geometry. HB02 found a quasi-periodic
bump and many step-like features purely due to the velocity field
of spiral waves. In their work, predicted profiles are obtained
for various spiral wave parameters, and these authors concluded
that the multiple sub-peaks are generally more prominent for
larger spiral wave numbers.

In our current paper we will focus our attention on issues (ii)
and (iii) where an X-ray source geometry and the velocity field
are both important ingredients for determination of the iron line.
Our model, being motivated by both HB02 and MM03, postulates a
localized X-ray flare via a relatively small magnetic reconnection
(small enough not to destroy the main structure) somewhere above
the innermost region of the accretion disk around a
rapidly-rotating black hole, in the presence of the spiral
accretion flow superposed onto the background Keplerian motion,
which is a different set-up from HB02. We also consider a
different height of the flare. The X-ray flare will then
illuminate the nearby disk material and produce cold iron
fluorescent emission line at 6.4 keV in the rest frame. The iron
is considered to be either neutral or only weakly ionized (i.e.,
Fe1-Fe16) in the case of a moderately weak flare. Thus, the
effective iron line profile in our model will be determined
primarily by both the {\it spiral velocity field} and the {\it
location of the flare}.

The structure of this paper is as follows. In \S{2} we will
introduce the assumptions made and establish our flare model in
the context of the thin-disk line model in the presence of the
spiral motion. The results are presented in \S{3} where we will
show our theoretical line profiles for various situations
including all the special/general relativistic effects, both from
the flare source to the disk and from the disk to the observer.
The apparent disk image (redshift and blackbody temperature) are
also presented. The discussion and concluding remarks are given in
the last section, \S{4}.

\section{Assumptions \& Basic Equations}
First, we construct a spiral motion, superimposed onto the
unperturbed Keplerian orbit, in such a way that it qualitatively
mimics to some degree the characteristic velocity field found in
H01 and MM03. Then we introduce the geometry of our postulated
flare X-ray source. For computing relativistic geodesics of
emitted photons from the iron to the observer, the ray-tracing
method is employed. All the general/special relativistic effects
(longitudinal Doppler shift, relativistic beaming effect, bending
of light, and the gravitational redshift) are included in our
computations, both from the flare to the disk and from the disk to
the flare.

\begin{figure}
    \epsscale{0.5}
    \plotone{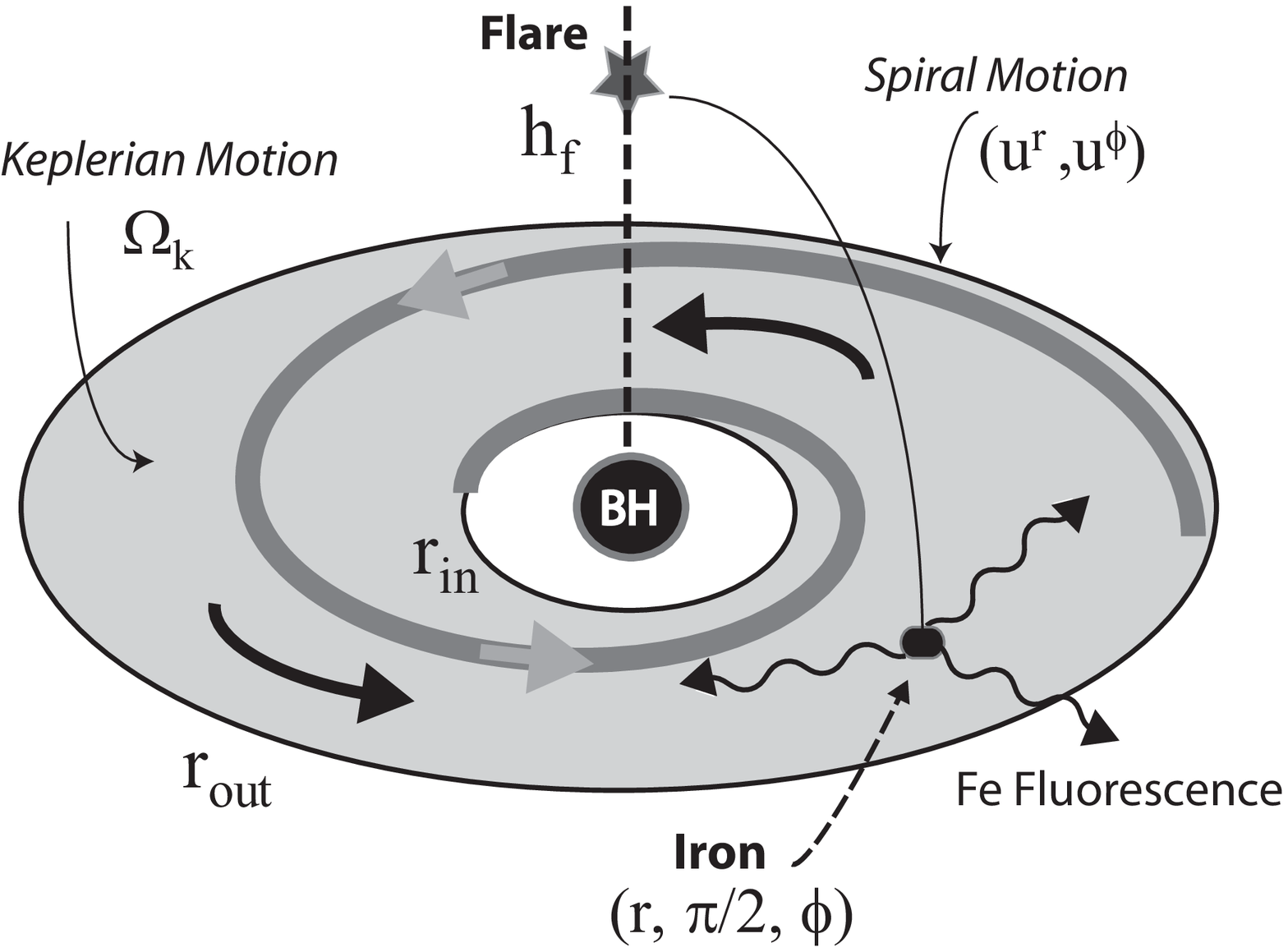}
    \caption{A cartoon of the the accretion disk in the presence of the spiral motion. The black hole is situated
             at the center above which an X-ray source (flare) is emitting primary photons. Not drawn to scale.}
    \label{fig:top}
\end{figure}

\placefigure{fig:top}

\subsection{Spiral Motion with the Keplerian Flow in the Disk}
Let us first assume that the space-time is stationary ($\partial /
\partial t=0$) and axially symmetric ($\partial /
\partial \phi =0$) in the equatorial plane ($\theta=\pi/2$).
Following the standard thin-disk models, the disk is considered to
be geometrically-thin ($h/r \sim 0.1$) and optically-thick
($\tau_{es}>1$), ranging from an inner radius $r_{in}$ to an outer
radius $r_{out}$. Here, $h$ is the scale-height of the thin disk,
and $\tau_{es}$ is the electron scattering optical depth. The
schematic geometry of our model is illustrated in
Figure~\ref{fig:top}. It shows a schematic view of the innermost
region along with a local flare at height $h_f$ above the disk
surface at ($r,\pi/2,\phi$). The physical meaning of the
illustrated notations will be explained in later sections.

The background geometry is described in the Boyer-Lindquist
coordinates as
\begin{eqnarray}
         ds^2 &=& -\left( 1-\frac{2Mr}{\Sigma} \right) dt^2
        - \frac{4aMr\sin^2\theta}{\Sigma} \,dt d\phi \nonumber \\
        & & + \frac{A\, \sin^2\theta}{\Sigma} \, d\phi^2
        + \frac{\Sigma}{\Delta}\, dr^2 + \Sigma\, d\theta^2 \ ,
\end{eqnarray}
where $\Delta \equiv r^2-2Mr+a^2,~\Sigma \equiv r^2+a^2 \cos^2,~A
\equiv (r^2+a^2)^2-a^2 \Delta \sin^2 \theta$, and $M$ and $a$ are
the black hole mass and its specific angular momentum,
respectively. The metric signature is $(-,+,+,+)$. Geometrized
units are used such that $G=c=1$ where $G$ and $c$ are
respectively the gravitational constant and the speed of light. In
this paper, distance is normalized by the gravitational radius
$r_g \equiv GM/c^2$. For instance, $r_g \sim 1.5 \times 10^{12}$
cm for $10^7 M_{\Sun}$ black hole mass. The unperturbed accretion
disk possesses the Keplerian accretion flow with the angular
velocity $\Omega_k = 1 / (r^{3/2}+a)$ where $r$ is the position of
the emitter (i.e., iron) in the disk surface. In order to discuss
physical quantities let us introduce a zero angular momentum
observer (ZAMO) in a locally non-rotating reference frame (LNRF)
which is a locally flat space. As seen by a ZAMO, the physical
three-velocity (radial and azimuthal components) of the
unperturbed Keplerian flow is described by
\begin{eqnarray}
v^{\hat{r}}_k &=& 0 \ , \\  v^{\hat{\phi}}_k &=& \frac{A}{\Sigma
\Delta^{1/2}} (\Omega_k -\omega) \ ,
\end{eqnarray}
where $\omega \equiv 2aMr/A$ is the angular velocity of the ZAMO
and the subscript {\it ``k''} denotes the Keplerian value and
$\hat{}$ ({\it hat}) represents quantities measured by the ZAMO.
We set $r_{out}=30r_g$ in our calculations for two reasons. It is
both because we are interested in the relativistic line broadening
in the inner region of the disk and because our attention is
focused on the broad redtail. The inner radius $r_{in}$ coincides
with the marginally stable circular orbit at $r_{ms}$ below which
the accreting flow starts spiralling-in towards the event horizon
at $r_h$ along geodesics (free-fall with $u^r \ne 0$). Since the
radial-gradient of the density within $r_{ms}$ (in the plunging
region) is very large due to the free-fall trajectory, the optical
depth $\tau_{es}$ there is not large enough to maintain the
optically-thick assumption. Thus, we do not consider the accreting
matter within $r_{ms}$ for fluorescence.

As suggested by H01 and MM03, the accretion disk is no longer
Keplerian in the presence of magnetic fields because the field
lines, which the accreting particles are frozen-in to, start
interacting with one another. The MRI under a differential
rotation is then expected to occur and trigger the perturbations
on the otherwise ordered Keplerian flow. As a result, the density
distribution starts appearing as an asymmetric, tightly wrapped
spiral pattern. There could be multiple-armed spiral waves, but
H01, MM03 and CT01 show that the spiral structure will eventually
evolve into {\it a single-armed pattern} after a certain rotation
period. For instance, in MM03 a one-armed spiral starts appearing
at $t \sim 10t_0$ where $t_0$ is one orbital time at $r \sim
50r_g$, while CT01 has found a single-armed spiral pattern after
about 200 orbits. Due to these findings by various previous
authors, we only consider a one-armed spiral wave here. In H01,
the specific angular momentum of the flow tends to fluctuate
around the Keplerian value due to the MRI, and therefore we will
phenomenologically include a similar effect caused by the spiral
perturbations in the velocity. In order to qualitatively reproduce
similar spiral motions found in their numerical results, we follow
HB02 and use a simple representation of the bulk velocity field
(radial and azimuthal components) of the spiral wave in the LNRF
as
\begin{eqnarray}
v^{\hat{r}}_{sw} &=& - A_r ~e^{-(r-r_{in}) / \Delta_{sw}}
\sin^{\gamma_o} \left[k_r (r-r_{in}) + m \phi /2 - \phi_{sw}/2 \right] \ , \label{eq:vr} \\
v^{\hat{\phi}}_{sw} &=& A_{\phi} ~e^{-(r-r_{in}) / \Delta_{sw}}
\sin \left[k_r (r-r_{in}) + m \phi - \phi_{sw} \right] \ ,
\label{eq:vphi}
\end{eqnarray}
where the subscript {\it ``sw''} denotes the spiral wave. Such a
spiral motion can qualitatively mimic the velocity perturbation
discovered by H01 and MM03, and hence it is not an arbitrary
choice. The characteristic structure of the spiral motion is thus
uniquely determined by a set of parameters $(A_r, A_{\phi}, m,
k_r, \Delta_{sw}, \gamma_o, \phi_{sw})$.

The number of parameters can be reduced by making reasonably
acceptable assumptions in the following way.  CT01 and MM03 both
show that the azimuthal wave number $m$ is fixed to be $m=1$ for a
single-armed spiral wave in a late stage of the spiral evolution.
$\gamma_o$ determines the width of the spiral pattern and turns
out to be ineffective to the qualitative features of the line
profile. Therefore, it is held constant at $\gamma_o=2$. Given
these fixed parameters, the actual free parameters are $(A_r,
A_{\phi}, k_r, \Delta_{sw}, \phi_{sw})$. MM03 shows that the wave
amplitude $A_r$ and $A_{\phi}$ are unlikely to be very large
($u^r$ only exceeds 0.1c). Besides, a large amplitude could
perturb and destroy the whole disk structure. Therefore the
amplitude is chosen to be relatively small ($A_r=A_{\phi}=0.1$) in
most of our calculations in \S{3.1}. However, we look for the
possible effects of amplitude variations. $k_r$ characterizes a
tightness (the number of winding) of the spiral pattern, and the
effective (radial) range of the spiral motion is controlled by
$\Delta_{sw}$. For instance, the spiral is more tightly packed
when $k_r$ is large, and it is more centrally (i.e., radially
inward) concentrated when $\Delta_{sw}$ is small. $\phi_{sw}$
denotes the phase of the spiral. $\phi_{sw}$ becomes an important
factor for determining the line profile in the phase-dependent
case. It is fixed otherwise. It is clear that the spiral motion is
non-axisymmetric and dependent on its phase $\phi_{sw}$. To avoid
the phase-dependence of the line profile, we hold the spiral phase
constant, leaving us the few free parameters ($A_r, A_{\phi}, k_r,
\Delta_{sw}$), for specifying a spiral motion. However, we will
later investigate the phase-evolved line profile holding every
parameter fixed except for $\phi_{sw}$. Note that $\phi=270\degr$
(equivalently $-90\degr$) coincides with the observer's azimuthal
angle, and $v^{\hat{r}}_{sw}<0$ for accretion. $v^{\hat{r}}$ and
$v^{\hat{\phi}}$ are set to be in-phase in equation~(\ref{eq:vr})
and equation~(\ref{eq:vphi}), which turns out to be unimportant to
the end results.

\begin{figure}
    \epsscale{0.4}
    \plotone{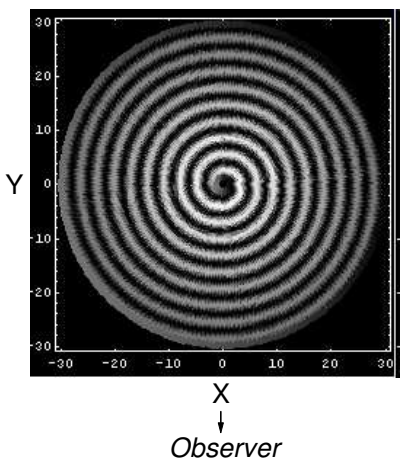}
    \caption{The spiral structure of the radial four-velocity component $u^r$
             with $k_r=1.0,~\Delta_{sw}=30r_g$ and $A_r=A_{\phi}=0.1$. The inner disk region ($-30r_g \le X \le 30 r_g$ and
             $-30 r_g \le Y \le 30 r_g$) is shown. A distant observer is situated in the azimuthal direction of $270\degr$.
             }
    \label{fig:spiral}
\end{figure}

\begin{figure}
    \epsscale{0.8}
    \plotone{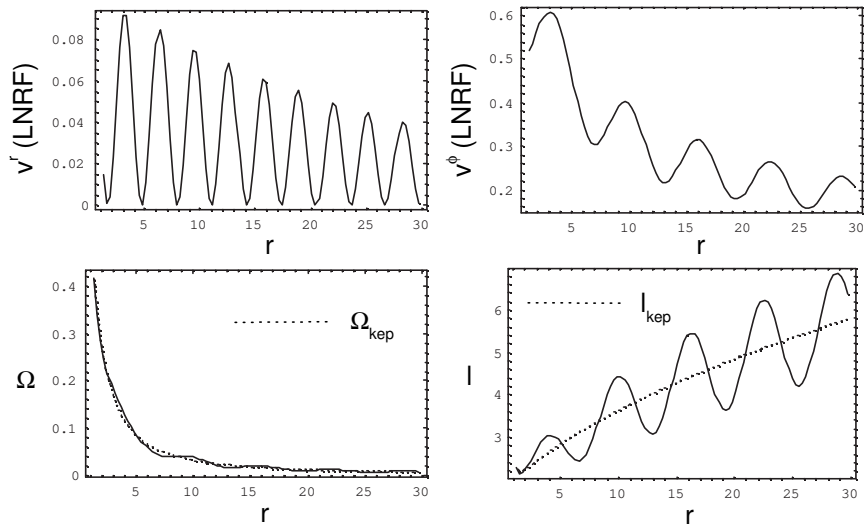}
    \caption{The velocity profiles corresponding to Figure~\ref{fig:spiral}. The radial three-velocity $v^{\hat{r}}$ (upper-left)
             in LNRF,
             the azimuthal three-velocity $v^{\hat{\phi}}$ (upper-right) in LNRF, the effective angular velocity $\Omega$ (lower-left)
             and the specific angular momentum $\ell$ (lower-right) are respectively displayed along the X-axis ($\phi=0\degr$)
             for $k_r=1.0,~\Delta_{sw}=30r_g$ and $A_r=A_{\phi}=0.1$.
             The {\it dotted curves} are for the unperturbed Keplerian orbit.}
    \label{fig:velocity}
\end{figure}

\placefigure{fig:spiral}
\placefigure{fig:velocity}

We assume that the effective accretion flow seen by a ZAMO in the
LNRF is described as the sum of these velocity fields; the spiral
orbit being superposed onto the unperturbed Keplerian orbit. By
adding each physical three-velocity field in the LNRF, we get
\begin{eqnarray}
v^{\hat{r}} &=& v^{\hat{r}}_k + v^{\hat{r}}_{sw} \ , \\
v^{\hat{\phi}} &=& v^{\hat{\phi}}_k + v^{\hat{\phi}}_{sw} \ .
\end{eqnarray}
Since $(v^{\hat{r}}, v^{\hat{\phi}})$ is not axisymmetric, the net
velocity field is also non-axisymmetric. The corresponding
four-velocity field of the effective flow is then written as
\begin{eqnarray}
(u^t,~u^r,~u^{\theta},~u^{\phi}) = u^t \left(1,
~\frac{\Delta}{A^{1/2}} v^{\hat{r}}, ~0 , ~\frac{\Sigma
\Delta^{1/2}}{A \sin\theta} v^{\hat{\phi}}+\omega \right),
\label{eq:4-velocity}
\end{eqnarray}
where $u^t = [-\hat{V}^2 \Delta \Sigma / A + 1-2r / \Sigma + A
\omega^2 / \Sigma]^{-1/2}$ and $\hat{V}^2 \equiv (v^{\hat{r}})^2 +
(v^{\hat{\phi}})^2$ is the square of the physical three-velocity
of the perturbed flow in the LNRF. $\omega \equiv 2Mar/A$ is the
angular velocity of the ZAMO with respect to a distant inertial
frame. Thus, the effective velocity field including the spiral
motion in Keplerian background orbit is determined once we specify
$v^{\hat{r}}_{sw}$ and $v^{\hat{\phi}}_{sw}$ through a set of free
parameters $(A_r, A_{\phi},k_r, \Delta_{sw})$.
Figure~\ref{fig:spiral} shows an example of a perturbed accreting
flow where the radial four-velocity component $u^r$ described by
equation~(\ref{eq:4-velocity}) is plotted in the equatorial plane
($-30 \le X \le 30$ and $-30 \le Y \le 30$). The adopted
parameters are $k_r=1.0,~\Delta_{sw}=30r_g$ for illustration
purpose. The corresponding velocity profile at $\phi=0\degr$
(along Y=0 in Figure~\ref{fig:spiral}) is shown in
Figure~\ref{fig:velocity}. From upper-left clockwise, the radial
three-velocity $v^{\hat{r}}$, the azimuthal three-velocity
$v^{\hat{\phi}}$, the effective angular velocity $\Omega$ and the
specific angular momentum of the gas $\ell$ are displayed,
respectively. For comparison, the dotted curves are also shown for
Keplerian motion.

\subsection{A Localized X-ray Flare Source}

In our model the X-ray flare source is a locally stationary active
region (e.g., flaring sites in corona) at some height $h_f$ on the
symmetry axis above the innermost disk. Following the practice
adopted in many standard disk-corona model for the Fe fluorescent
lines, we also approximate such a source as a point-like X-ray
source on the black hole symmetry axis (see Figure~\ref{fig:top}).
This assumption should be justified because the hottest region
should be above the innermost disk because the multi-color disk
temperature scales as $\propto r^{-3/4}$. That should be very
close to the rotation axis of the hole. We also assume that the
entire disk is neutral or at most only weakly ionized by such a
radiation source, producing a neutral-like fluorescence at $6.4$
keV in the source frame. In this manner we can investigate the
effects exclusively due to the spiral wave alone, by comparing our
results with those obtained for the standard Keplerian disk model
without spiral waves.

\subsection{Relativistic Disk Emissivity}
In Newtonian geometry, specifying the height of the X-ray source
$h_f$ would simply allow us to compute the emissivity law of the
illuminated disk. In addition, it is also important, especially in
the innermost region in the relativistic disk, to take into
account the general relativistic bending effect (either redshift
or blueshift) on the photons from the flare to the disk. Due to
this effect, the intensity of the illuminating photon flux at the
disk should be subject to a significant deviation from the
Newtonian case. \cite{RB97} calculated this effect for their iron
lines predicted for Schwarzschild geometry, whereas
\cite{Martocchia96} calculated the photon trajectories from an
on-axis source to the underlying accretion disk in Kerr geometry.
We adopt and modify the work of the latter authors who found the
redshift factor $g_{sd}$ measured in a local disk frame (i.e.,
ZAMO) as
\begin{eqnarray}
g_{sd} =
\sqrt{\frac{(r^2+a^2-2r)(r^2+a^2+2a^2/r)}{(1-\hat{V}^2)(r^2+a^2-2r)(h^2_f+a^2)}}
, \
\end{eqnarray}
where we adopt the extreme Kerr parameter $a=0.998$ according to
recent theoretical/observational speculations that a black hole is
likely to rotate rapidly in some AGNs
\citep{Iwasawa96a,Iwasawa96b,Young98,Fab02,Wilms01}. Given the
canonical photon power-law index ($\Gamma_{PL} \sim 1.9$) for
Seyfert 1s \citep{Nandra94}, the actual local emissivity is
weighted by the factor of $g^{1.9}_{sd}$, which yields the net
local axisymmetric emissivity as
\begin{eqnarray}
\tilde{\epsilon}_d = \epsilon_d ~g^{1.9}_{sd} \ ,
\end{eqnarray}
where $\epsilon_d$ is the local Newtonian emissivity of the disk
expressed as
\begin{eqnarray}
\epsilon_d = \frac{h_f}{\left(r^2 + h_f^2 \right)^{3/2}}  \ .
\end{eqnarray}
Here, $r$ is the radial position of the emitter (iron) in the
disk.

\subsection{Photon Trajectories}
When fluorescent photons emitted from the disk reach a distant
observer or telescope, the following geodesic equation in the
integral form must be satisfied \citep[]{carter68,chandra83}:
\begin{eqnarray}
\int_{r_{em}}^{r_{obs}} {{1 \over \sqrt{R(r)}}\,dr} = \pm
\int_{\theta_{em}}^{\theta_{obs}} {{1 \over
\sqrt{\Theta(\theta)}}\,d\theta} \ ,  \label{eq:geo}
\end{eqnarray}
where
\begin{eqnarray}
 R(r,\lambda,Q) &=&  r^4+(\lambda^2-Q)^2+2 (Q+\lambda^2)
 r - Q, \label{eq:radial} \\
 \Theta(\theta,\lambda,Q)  &=&  Q -(\lambda
 \cdot cot\theta)^2,  \label{eq:angular}
\end{eqnarray}
Here an observer's distance $r_{obs}$ and its inclination angle
$\theta_{obs}$ need to be specified. $(r_{em}, \theta_{em})$ is
the emitter's (i.e., iron) position on the disk. For a distant
observer, we take $r_{obs}=\infty$. $\lambda$ and $Q$ are two
constants of motion along a geodesic, which are closely related to
the axial component of the angular momentum of photons
\cite[]{carter68}. We will make full use of the elliptic integrals
to numerically evaluate equation~(\ref{eq:geo}) in the ray-tracing
approach \citep[]{cadez98,fanton97} for efficient computations. We
will first search for an observable photon emitted from $r_{em}$
in the disk, and then calculate the corresponding redshift
$g_{do}$. Given the 4-velocity of the perturbed flow (iron) and
the 4-momentum of the null geodesic (photon), we obtain the
redshift by
\begin{eqnarray}
g_{do} \equiv \frac{E_{obs}} {E_{em}} = 1/[u^t(1-\lambda \Omega -
v^{\hat{r}} \sqrt{R} /r^2)] \ ,
\end{eqnarray}
where the observer at a distant location $r_{obs}$ is stationary.
$E_{em}$ and $E_{obs}$ are the local photon energy and its
observed energy, respectively. We consider a photographic plate of
the observer's (or telescope's) window at $r_{obs}$ facing
straight the accretion disk with the inclination angle
$\theta_{obs}$. Images of the disk are projected onto this window
with ($\alpha, \beta$) where $\alpha,\beta$ are the impact
parameters of the observed photons in observer's sky. In our
calculations, the window roughly contains $280 \times 280$ pixels
(corresponding to the spatial resolution of $\sim 0.23 r_g$) in
which the images of the disk are generated in false color. We only
consider the observed photons normal to this window.

\subsection{Predicted Fluorescent Emission Lines}
Once we know the redshift factor $g_{do}$ and the surface
emissivity of the disk $\epsilon_d$ including the bending effect
expressed by $g_{sd}$, the observed photon flux is obtained by
\begin{eqnarray}
F_{obs}(E_{obs}) = \int_{source} g_{do}^4 ~I_{em} ~d\Omega,
\label{eq:flux}
\end{eqnarray}
where d$\Omega$ is the solid angle subtended by the disk in the
observer's frame. Here, the local intensity is approximated by the
$\delta$-function of the monochromatic photon energy $E_{em}$ in
the source frame as $I_{em}=\tilde{\epsilon}_d \cdot
\delta(E_{obs}-E_{em})$. Broadening of the line photons due to
turbulence may be insignificant \citep{Fab00}. The local emitting
region is assumed to be optically thick in the absence of a thick
electron-scattering atmosphere above the disk so that the
limb-darkening effect is unimportant, especially for a small
inclination angle $\theta_{obs} $\citep{Chen91,Laor91,Bao94}. In a
statistical sense, it seems to be favored that the averaged
$\theta_{obs}$ for Seyfert 1s is around $\sim 30\degr$
\citep[]{Antonucchi85,Schmitt01}. Therefore, we adopt
$\theta_{obs}=30\degr$ throughout this paper. Notice that the
additional redshift factor $g_{sd}$ due to the photon bending
(from the source to the disk plane) is included via our emissivity
prescription $\tilde{\epsilon}_d$ in the modified local intensity
$I_{em}$.

Using the obtained fluorescent flux, we can estimate the
equivalent blackbody temperature measured by a distant observer
$T_{obs}$ from the Stefan-Boltzmann law
\begin{eqnarray}
T_{obs} = ( F_{obs}/\sigma_{SB} )^{1/4} \ ,
\end{eqnarray}
which is a function of the redshift factor $g_{do}$, the position
of the emitter in the disk $(r, \pi/2, \phi)$ for a given height
of the flare $h_f$. $T_{obs}$ will allow us to see the temperature
distribution in the perturbed disk. If the spiral wave is fixed in
the rotating disk frame, the line profile becomes phase-dependent
of $\phi_{sw}$, and its characteristic rotational period should be
the Keplerian value.

\section{Results}
Since there are a few degrees of freedom in our parameter space
that primarily characterize the line shape, we need to carefully
and systematically examine the effects of each parameter alone.
First, we will see the effects of various spiral patterns by
varying ($k_r, ~\Delta_{sw}$) for a given spiral phase
$\phi_{sw}=0\degr$. Later in \S{3.1.2}, the spiral phase
$\phi_{sw}$ is varied under the assumption that the spiral
structure is fixed in the rotating disk frame. Our primary goal in
the present paper is to demonstrate any detectable features seen
in the iron line exclusively caused by a spiral perturbed velocity
field, but for comparison purpose let us consider two different
source height: $h_f=4 r_g$ (a flare close to the hole) and $h_f=10
r_g$ (a relatively distant flare). The selected parameters for various
models are tabulated in Table~\ref{model}. In all cases we set
$A_r=A_{\phi}$ (equal amplitude) unless otherwise stated.

\placetable{tab:model}

\subsection{Predicted Iron Line Profile}
Using the ray-tracing method, we are able to find roughly $\sim 6
\times 10^4$ photon trajectories between the accretion disk and
the observer. In the subsequent binning process for various photon
energy, we set the spectral resolution to be $E / \Delta E \sim
600$, equivalent to $\sim 10$ eV, which can be achieved with X-ray
Spectrometer (XRS) on board {\it Astro-E2}. We will first show the
dependence of the wave pattern ($A_r, A_{\phi}, k_r, \Delta_{sw}$)
in \S{3.1.1}. Lastly, in \S{3.1.2} we will explain the
phase-dependence of the profile.

\placefigure{fig:profile-1-1}
\placefigure{fig:profile-1-2}

\subsubsection{Effects of Various Spiral Motions}
\begin{figure}
    \epsscale{0.7}
    \plotone{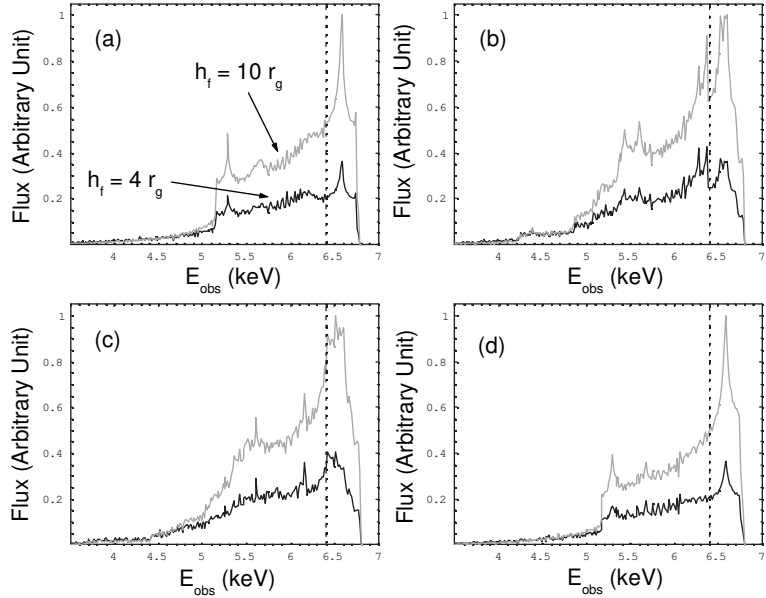}
    \caption{The predicted iron line profiles for various $k_r$ with $\Delta_{sw}=30 r_g$.
             $k_r=0.4, ~1.0, ~1.5$ and $0.4$ for (a), (b), (c) and (d), respectively. $A_r=A_{\phi}=0.1$ in all cases but (d)
             where $A_r=0.1 A_{\phi}=0.01$. A different source height is considered: a nearby-source ($h_f=4
             r_g$) in the {\it dark curve} and a distant source ($h_f=10 r_g$) in the {\it light grey
             curve}. The spectral resolution $E / \Delta E$ is
             $\sim 600$ (corresponding to $\sim 10$ ev) achievable with {\it Astro-E2} XRS.
             The {\it vertical dotted line} denotes 6.4 keV line in the rest
             frame.}
    \label{fig:profile-1-1}
\end{figure}

\begin{figure}
    \epsscale{0.7}
    \plotone{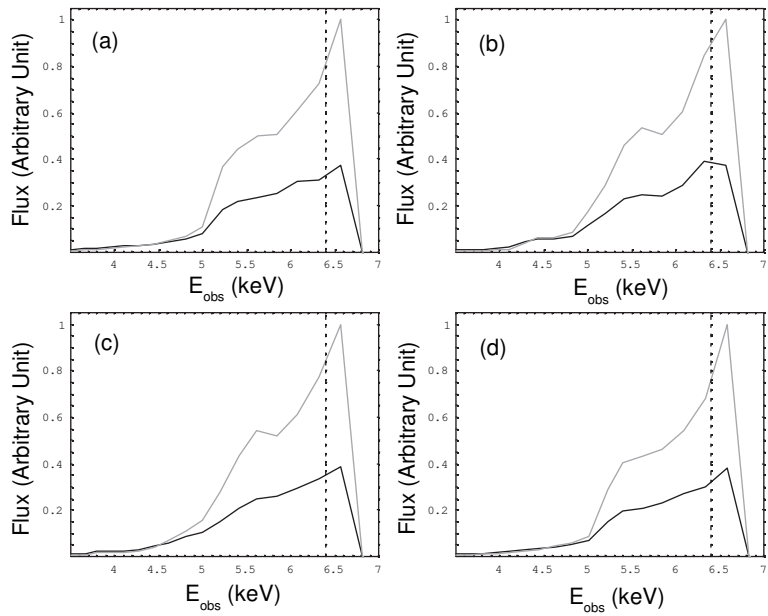}
    \caption{The predicted iron line profiles with the same parameter sets as in Figure~\ref{fig:profile-1-1}
             except that the spectral
             resolution $E / \Delta E$ is $\sim 50$ (corresponding to $\sim 120$ eV) already achieved with {\it ASCA}. }
    \label{fig:profile-1-2}
\end{figure}

Figure~\ref{fig:profile-1-1} shows the predicted line profiles for
various $k_r$ with $\Delta_{sw} =30r_g$. The horizontal axis shows
the observed photon energy in keV while the vertical axis denotes
the observed photon flux in an arbitrary unit. $k_r=0.4,~1.0$ and
$1.5$ for (a), (b) and (c), while in (d) $k_r=0.4$ but $A_r=0.1
A_{\phi}=0.01$ in order to see $v^{\hat{r}}$-dependence. The dark
solid curve represents the case where $h_f=4 r_g$ while the light
grey curve denotes the case with $h_f=10 r_g$. The vertical dotted
line shows a cold fluorescent line at 6.4 keV in the rest frame.
In general, randomly formed sub-peaks are present due to the
spiral velocity field, consistent with HB02. We further discover
that there are normally two types of peaks existing in the
profile: (i) relatively sharp sub-peaks and (ii) very spiky
quasi-periodic multiple-peaks. The two kinds of peaks are in
general simultaneously formed and are distinguishable in many
cases, which is a new result inherent to our model. The spike-like
features (ii) seem to be more prominent as $k_r$ increases because
the winding of the spiral arm is amplified although the sharp
sub-peaks (i) are less obvious. We can clearly observe an unusual
double-peak structure (or a quasi-triple-peak structure) in (a).
As $k_r$ becomes larger, quasi-periodic spike-like peaks tend to
dominate over the sharp peaks, smoothing out the (classical)
prominent peaks. When the spiral wave is tightly packed
($k_r=1.5$), the profile seems to become almost monotonic in the
sense that the sharp sub-peak structure described as (i) is no
longer clearly formed and instead be dominated by spiky
multiple-peaks classified as (ii). This is justified from another
model with $k_r=2.0$ (not shown here). The radial velocity
component $v^{\hat{r}}$ appears to be unimportant (or relatively
insignificant) in the qualitative formation of the randomly
distributed peaks. The flux with a distant X-ray source ($h_f=10
r_g$) is larger than that with a nearby source ($h_f=4 r_g$) at
most of the energy $E_{obs}$ band. This can be understood by the
fact that for $h_f=4 r_g$ the emissivity form $\tilde{\epsilon}_d$
becomes extremely centrally-concentrated because of a huge effect
of light bending via $g_{sd}$, which yields a very steep
emissivity in the innermost region. In this case, the innermost
gas with a high redshift factor (hence emitting low energy
photons) plays an important role in producing the iron line, which
has almost no significant contributions to high energy fluorescent
photons. In the case of $h_f=10 r_g$, on the other hand, the
bending of light is not as crucial as the previous case, and
therefore such a relatively distant X-ray source can illuminate
the whole disk (gas) in an even manner to produce more flux of
blueshifted photons.

To better illustrate an observational difference associated with a
currently available spectral resolution, the same line profiles
are obtained with $E/\Delta E = 50$ (corresponding to $\sim 120$
eV already achieved with {\it ASCA}) in
Figure~\ref{fig:profile-1-2} where for comparison the same
parameter sets are adopted as in Figure~\ref{fig:profile-1-1}.
With such an energy resolution, it seems very difficult to
differentiate one model from the other (for instance, the one with
low $k_r$-value and the one with high $k_r$-value) and even
impossible to distinguish the present model (with perturbations)
from the standard disk-line model. This, however, will be
distinguishable observationally with a much better spectral
resolving power that should be available in the next generation
X-ray telescopes such as {\it Astro-E2}.

\begin{figure}
    \epsscale{0.7}
    \plotone{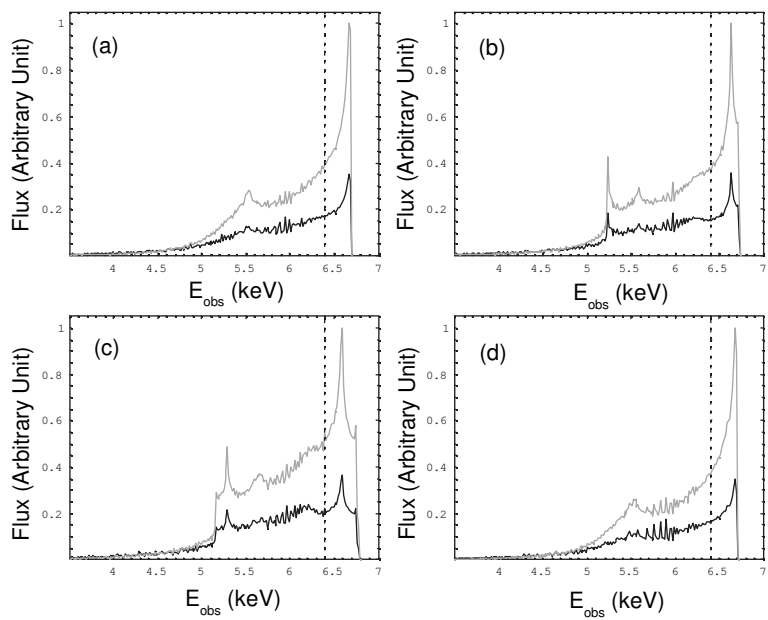}
    \caption{The predicted iron line profiles for various $\Delta_{sw}$ with $k_r=0.4$.
             $\Delta_{sw}/r_g=5, ~15, ~30$ and $5$ for (a), (b), (c) and (d), respectively. $A_r=A_{\phi}=0.1$ in all cases
             but (d) where
             $A_r=0.1 A_{\phi}=0.01$. }
    \label{fig:profile-2}
\end{figure}

\placefigure{fig:profile-2}

Figure~\ref{fig:profile-2} displays the profiles for various
$\Delta_{sw}$ with $k_r=0.4$ where $\Delta_{sw}$ determines the
effective (radial) distance of the spiral perturbation. We use
$\Delta_{sw}/r_g=5,~15$ and $30$ for (a), (b) and (c),
respectively, while in (d) we adopt $\Delta_{sw}/r_g=5$ and
$A_r=0.1 A_{\phi}=0.01$. The degree of the spiky multiple-peak
looks roughly unchanged in every case regardless of $\Delta_{sw}$,
and those multiple-peaks are already present even when the spiral
is centrally concentrated in (a). However, the overall line shape
in (a) is similar to that of the standard model in terms of the
double-peak structure (the third peak between the red one and the
blue one is almost unseen). Therefore, it may be difficult to
detect significant spiral effects when the perturbation is
concentrated radially-inward. Again, the radial component of the
spiral motion appears to be unimportant in this case, too.

\begin{figure}
    \epsscale{0.7}
    \plotone{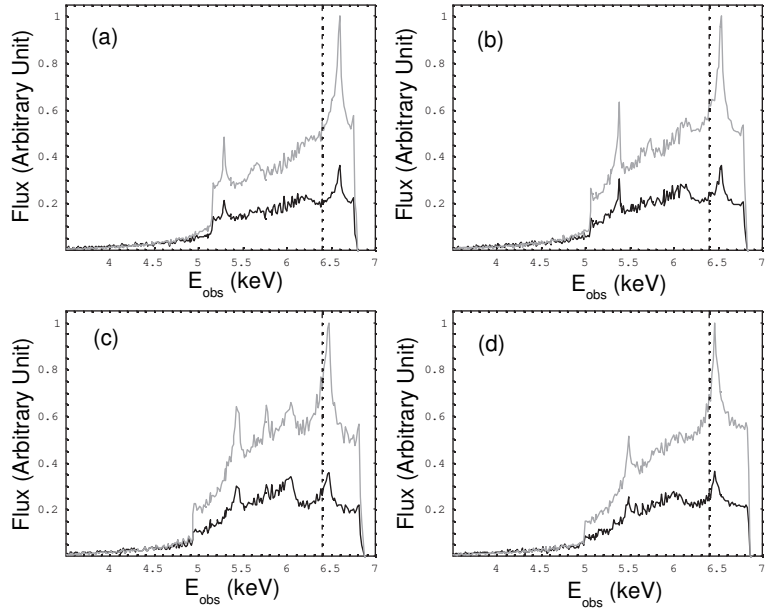}
    \caption{The predicted iron line profiles for various amplitude ($A_r,A_{\phi}$) with $k_r=0.4$ and $\Delta_{sw}=30r_g$.
             $A_r=A_{\phi}=0.1,~0.15$ and $0.2$ for (a), (b) and (c), respectively. In (d), $A_r=0.1 A_{\phi}=0.02$. }
    \label{fig:profile-3}
\end{figure}

\placefigure{fig:profile-3}

As we earlier mentioned, more effects of the spiral motion is
intuitively expected when the amplitude is large. In
Figure~\ref{fig:profile-3}, $A_r=A_{\phi}=0.1,~0.15$ and $0.2$ in
(a), (b) and (c), respectively while in (d) $A_r=0.1
A_{\phi}=0.02$. As expected, a large amplitude tends to
complicates the line shape more effectively. Specifically, the
number of sharp sub-peaks appears to increase whereas the degree
of the spiky multiple-peaks remains almost unchanged when the
amplitude increases. For example, the number of sub-peaks can be
as high as $\sim 6$ when $A_r=0.2$ in (c) while it is $\sim 3$
when $A_r=0.1$ in (a). For a small radial component in (d), there
seems to be no significant difference from (a).

From what we find so far, it seems reasonable to say that the
resulting line profile can be made very different from the
classical one basically in two ways: (1) a tightly packed spiral
wave with large $k_r$ tends to produce more spiky multiple-peaks
(wiggling) or (2) a large azimuthal amplitude $A_{\phi}$ tends to
produce more sharp sub-peaks, unless the spiral wave is restricted
to the very inner region (small $\Delta_{sw}$).

\placefigure{fig:phase}
\placefigure{fig:intensity}

\subsubsection{Phase-Dependence }
\begin{figure}
    \epsscale{0.8}
    \plotone{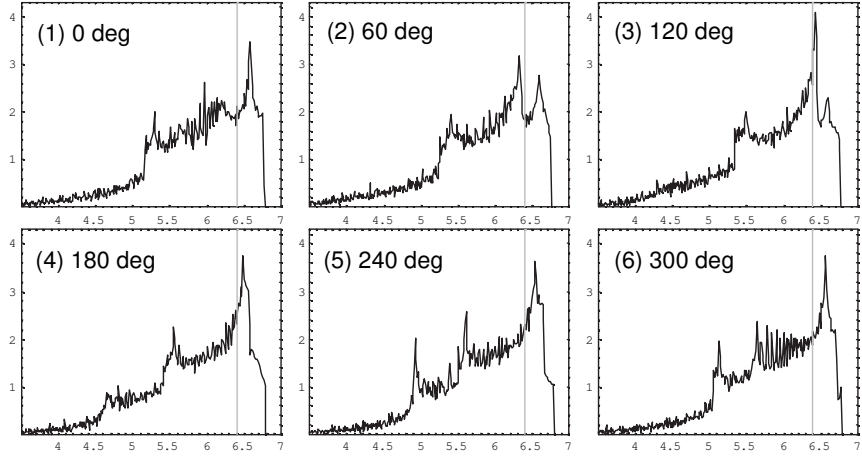}
    \caption{The phase-evolved iron line profiles from a spiral wave with $k_r=0.4,~\Delta_{sw}=30r_g$ and $A_r=A_{\phi}=0.1$.
             The source height is at $h_f=4 r_g$. The phase of the spiral wave $\phi_{sw}$ is progressing from (1)~$0\degr$
             to (6)~$300\degr$ by $60\degr$. The horizontal axis is the observed
             photon energy $E_{obs}$ (keV) while the
             vertical axis shows the observed photon flux in arbitrary units. }
    \label{fig:phase}
\end{figure}

\begin{figure}
    \epsscale{0.6}
    \plotone{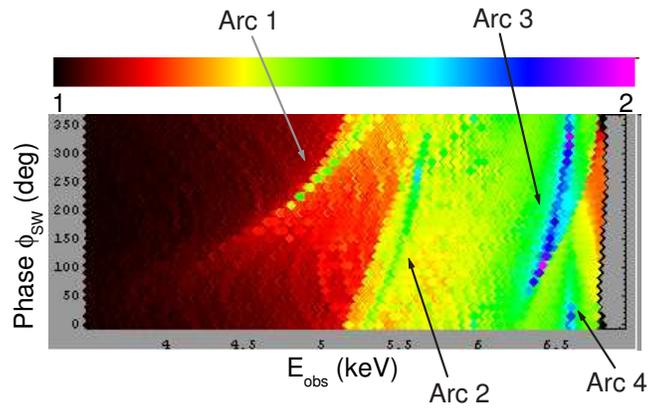}
    \caption{The intensity map of the global evolution of the iron line in the energy-phase $E_{obs}-\phi_{sw}$ space
             corresponding to Figure~\ref{fig:phase} with the same
             parameter set being used.
             The phase of the spiral wave $\phi_{sw}$ is progressing from $0\degr$ to $345\degr$ by
             $15\degr$.The intensity is normalized between 1 and
             2.
             [{\it A color version of this figure is available at the electronic
             edition of the Journal.}] }
    \label{fig:intensity}
\end{figure}

In Figure~\ref{fig:phase} we show the phase-dependence of the line
profiles for a fixed spiral structure, with
$k_r=0.4,~\Delta_{sw}=30 r_g$ and $A_r=A_{\phi}=0.1$, varying
$\phi_{sw}$ from $0\degr$ to $300\degr$ by $60\degr$. The source
height is $h_f=4 r_g$. The spiral wave is now fixed to the
rotating disk frame. In other words, the spiral is corotating with
the accreting gas. The distant observer at $r_{obs} \sim \infty$
is situated at $270\degr$ (or equivalently $-90\degr$) in its
azimuthal position.

First of all, the profile appears to be constantly broadened over
the photon energy $E_{obs}$ ($\gtrsim$ 4 keV) for the entire phase
of the spiral wave $\phi_{sw}$ probably due to a small radius of
the marginally stable orbit $r_{ms}$. The line shape also normally
contains a triple-peak structure (left-end peak, middle-peak and
right-end peak) with the dominating blue peak (or right-end peak)
at all the times. It can be classified as belonging to roughly two
categories depending on its characteristic shape: (i) the left-end
peak above 5 keV such (1), (2), (3) and (6) and (ii) the left-end
peak below 5 keV such as (4) and (5). The line energy of each peak
then shifts towards either higher energy or lower energy depending
on the phase $\phi_{sw}$ (i.e., peak transition). This trend can
be viewed in relation to the spiral phase $\phi_{sw}$ in the
following. The main cause of this complicated variation primarily
originates from the fact that some parts of the accretion disk is
more redshifted while the other parts, on the other hand, are more
blueshifted depending on the phase $\phi_{sw}$. Hence,
non-axisymmetry in the net velocity field is directly responsible
for producing such a variability for a fixed X-ray source $h_f$.
As discussed earlier, a choice of nearby source (smaller $h_f$)
tends to generate more frequent sub-peaks due to more exclusive
contribution from the innermost gas where the spiral perturbation
is greater. On the other hand, spiky quasi-periodic peaks do not
seem to change very much over the phase $\phi_{sw}$. We have also
confirmed that the above variation (or the transition of peaks)
with phase $\phi_{sw}$ is not so obvious when $k_r$ is large
because the velocity field becomes somewhat quasi-axisymmetric
(that is, the perturbation becomes more or less uniform
everywhere). Hence, the noticeable line variability would be more
expected when the perturbing spiral wave is not so tightly packed
(i.e., small $k_r$). The corresponding global variation with
$\phi_{sw}$ can also be seen in Figure~\ref{fig:intensity} where
the normalized line intensity is plotted as a function of the
observed photon energy $E_{obs}$ (in keV) and the spiral phase
$\phi_{sw}$ (in degrees), corresponding to Figure~\ref{fig:phase}.
[{\it A color version of this figure is available at the
electronic edition of the Journal.}] The intensity is normalized
between 1 (its minimum) and 2 (its maximum). An interesting
pattern here is the existence of four distinct ``arcs (i.e., the
transition of peaks)" present at different energy $E_{obs}$: arc 1
($E_{obs} \sim 4.5-5.3$ keV), arc 2 (5.3-5.7 keV), arc 3 (6.3-6.5
keV) and arc 4 ($\sim$ 6.6 keV), respectively. They individually
correspond to the sharp peaks seen in the Figure~\ref{fig:phase}.
Besides such arcs, a randomly-distributed, wiggling variation of
the line intensity is clearly present, especially at $E_{obs} \sim
6$ keV, due to the perturbed velocity field. This corresponds to
the quasi-periodic multiple-peaks in Figure~\ref{fig:phase}.

\placefigure{fig:map}

\subsection{Redshift \& Temperature Distribution}
\begin{figure}
    \epsscale{0.7}
    \plotone{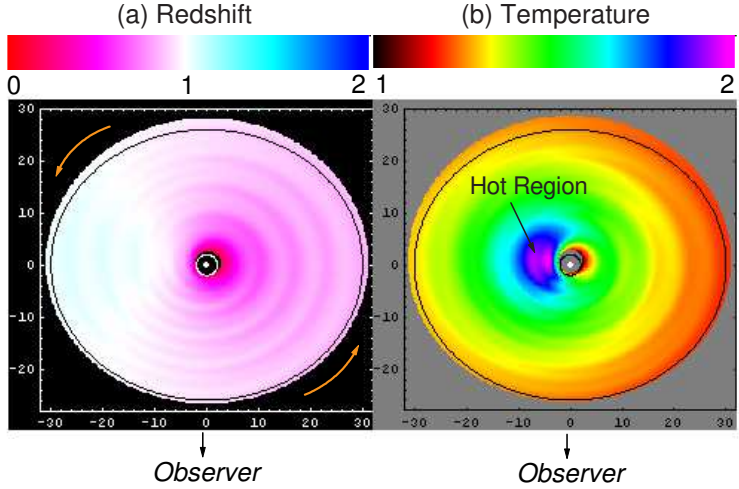}
    \caption{The redshift distribution $g_{do}$ of the emitted photons in the disk (left panel) and the equivalent blackbody temperature
             distribution $T_{obs}$ of the disk (right panel) for $k_r=1.0,~\Delta_{sw}=30 r_g$, and
             $A_r=A_{\phi}=0.1$. The phase of the spiral is $\phi_{sw}=0$ and the source is at $h_f=4 r_g$. A distant
             observer is situated in the direction of $270\degr$ (or equivalently $-90\degr$).
             The redshift $g_{do}$ of the observed photons is scaled
             between 0 and 2 in the left panel while the apparent
             disk temperature $T_{obs}$ is normalized between 1
             and 2 in the right panel.
             The {\it inner circle} denotes the event horizon $r_h$
             while the {\it outer circle} denotes the Newtonian circle
             with the radius of $30r_g$. The {\it arrow} in the left panel indicates
             the rotational direction of the disk. A distant
             observer is situated in the illustrated azimuthal
             position.
             [{\it A color version of this figure is available at the electronic
             edition of the Journal.}]}
    \label{fig:map}
\end{figure}

Assuming the blackbody emission from the fluorescent photons, we
estimate the equivalent temperature $T_{obs}$ measured by a
distant observer. Figure~\ref{fig:map} displays the spatial
distribution of the redshift $g_{do}$ and the temperature
$T_{obs}$ of the accreting gas for $k_r=1.0,~\Delta_{sw}=30
r_g,~A_r=A_{\phi}=0.1$ and $\phi_{sw}=0\degr$ where the redshift
$g_{do}$ is scaled between 0 (maximum redshift) and 2 (maximum
blueshift) whereas the temperature $T_{obs}$ is normalized by 1
(minimum temperature) and 2 (maximum temperature), respectively.
The source height is $h_f=10 r_g$ in this case. [{\it A color
version of this figure is available at the electronic edition of
the Journal.}] In the left panel the redshift factor $g_{do}$ is
projected onto the observer's local plane ($\alpha,~\beta$). The
outer circle and the inner circle denote the Newtonian radius of
$30r_g$ and the event horizon at $r_h$, respectively. The central
dot is the origin $(0,0)$ of the coordinate system. The observer
is situated at the azimuthal position of $270\degr$ (or
$-90\degr$). The spiral structure is clearly seen in both panels
along the spiral wave. In this figure, the accreting gas rotates
counterclockwise as indicated by the arrow. The receding part of
the disk ($-90\degr \le \phi \le 90\degr$) is relatively more
redshifted (darker) due to the classical Doppler motion. As $k_r$
increases, inhomogeneity of the redshift becomes more apparent. In
the right panel the apparent blackbody temperature $T_{obs}$ is
shown for the same parameter set. Again the spiral pattern is
present due to $T_{obs} \propto g_{do}$. The inner edge of the
disk shows a rapid temperature drop because of the extreme
gravitational Doppler redshift. The approaching side appears to be
hotter than the other regions due mainly to the classical Doppler
blueshift even under the spiral perturbation. Such a hot region
appears to be spatially narrow forming an interesting
two-dimensional shape (``hot spot'') in the approaching side of
the disk (this can be better seen in the color version of the
Figure~\ref{fig:map}). Interestingly, the spatial size of this
``hot spot'' is variable as the spiral phase $\phi_{sw}$
progresses whereas in the unperturbed disk it is constant.
Although we do not show the exact variable temperature map here,
it is found that the effective area of the ``hot spot'' appears to
vary periodically as the spiral wave rotates with the accreting
gas, which is due to the perturbed velocity field (and thus the
perturbed redshift distribution). The variation is found to be
larger when the spiral wave is mildly wounded (i.e., small $k_r$).
In the case of $h_f=10 r_g$, we also find a similar periodicity
except that the ``hot spot'' appears further out. This may imply
that such a variation can be manifested as a quasi-periodic change
in the observed photon flux with the Keplerian frequency of
$\Omega_k/(2\pi) \sim 4 \times 10^{-4}$ Hz at $r \sim 6 r_g$.

\section{Discussion and Concluding Remarks}

We modified and extended the work of HB02 on the iron lines under
spiral perturbing waves, by carrying out fully relativistic (both
special and general) calculations in the Kerr geometry. We
furthermore assume realistically that the net velocity field of
the accreting gas should be the sum of Keplerian motion and the
perturbing orbit in our model. The presence of spiral waves in the
magnetized accretion disk has been suggested already by various
previous numerical simulations (e.g., CT01, H01 and MM03). The
primary motivation of our current work is based on the numerical
results by these previous authors, which indicate that one-armed
spiral velocity field may be produced at a late stage of the disk
evolution due to the MRI for a differentially rotating disk. We
carefully chose the spiral velocity field in such a way that it
does qualitatively mimic the velocity perturbation already found
by H01 and MM03.

Our results in many respects confirms the work of HB02. However,
there are some important differences due to new results. Our
present work is new and original in various ways. For instance, we
extended the work of HB02 under an approximate Schwarzschild
geometry, to a full Kerr geometry. As a consequence, first of all,
our model generally allows for a {\it very long redtail} (below 4
keV), whereas the lines by HB02 extend at most only down to 4.5
keV. This is clearly due to our adoption of the Kerr geometry
which allows a smaller radius of marginal stability $r_{ms}$. In
terms of comparison with observations, our model is more realistic
than that of HB02 since observed lines sometimes exhibit such very
long redtail.

Moreover, we included {\it general relativistic bending of
photons, both from the X-ray source to the disk and from the disk
to the observer}. Bending of photons from the source to the disk
enables us to account for a more realistic centrally-concentrated
emissivity profile. Because of this effect, the net intensity of
the photon flux is altered from the Newtonian case. On the other
hand, HB02 does not include this bending effect although a similar
source geometry is assumed here.

We confirm that quasi-periodic peaks in the profile due to the
spiral motion in the accreting gas, which was reported by HB02 for
an approximate Schwarzschild geometry, are exhibited in the Kerr
geometry also. In addition, we newly find that the standard iron
line (meaning a broadened line with a double-peak structure) could
be dramatically modified depending on the type and/or degree of
such a spiral perturbation. Our results show that: (i) the profile
may commonly exhibits a multiple-peak (or non-standard
double-peak) even under a relatively small spiral perturbation,
(ii) the produced peaks can be divided into two kinds: sharp
sub-peaks and spiky multiple-peaks, (iii) the profile may possess
many spiky multiple-peaks for a tightly packed spiral wave, (iv) a
larger amplitude of the perturbation may produce more sharp
sub-peaks rather than spiky multiple-peaks, (v) the effect of the
spiral may not be significantly large enough for detection when
the perturbation is centrally concentrated, and from the
phase-evolution we have learned that (vi) the characteristic line
shape (i.e., relative position of the line peaks) seems to be very
sensitive to the azimuthal phase of the spiral wave, especially
for large $k_r$ and/or a small $h_f$.

The result (i) is qualitatively similar to what HB02 has found.
However, in addition we find more new features in our results (ii)
through (v) for various spiral patterns. We further find clearer
line variability (e.g., peak transition) in (vi) assuming that the
spiral wave is fixed in the rotating disk frame, although this is
not so obvious in HB02. If such a periodicity is found from future
observations, it may be attributed to the presence of such a
spiral rotation.

Finally, one of our goals is to find any possible observable
signatures due to the spiral motion imprinted in the iron
fluorescent line. It is beyond the sensitivity of the current
X-ray missions, such as {\it Chandra} and {\it XMM-Newton}, to
detect some of our predicted features, such as the sharp sub-peaks
and/or very narrow multiple-peaks. However, it will be
observationally feasible to detect such signatures by future X-ray
missions - if the spiral wave is actually involved in the
production of fluorescent emission line. For instance, we choose
the spectral resolving power to be $E / \Delta E \sim 600$
(equivalent to $\sim$ 10 eV) in all the computations, which can be
reached by next generation X-ray missions. For instance, Japan/USA
{\it Astro-E2}, scheduled to be launched in 2005, should be able
to achieve $E / \Delta E\sim 500-700$ (equivalent to the spectral
resolution of 10-13 eV or 6 eV of FWHM) around 6.4 keV, and NASA's
{\it Constellation-X} is designed to accomplish $E /\Delta E \sim
300-1500$. Also, the next European X-ray satellite, {\it XEUS}, a
potential successor to {\it XMM-Newton}, will have $E / \Delta E
\sim 1000$ at 6 keV. These energy resolutions should be more than
sufficient to identify the predicted multiple sub-peaks in the
profile. Therefore, {\it our model can be tested with these future
observations} \citep[e.g.][]{Reynolds03}.  Then it will allow us
to evaluate whether our proposed model (perturbed accreting gas)
can correctly describe the observed line profiles. If so, further,
careful comparison with observation may enable us to narrow down
some of the parameters for the spiral wave structure.

Previous standard disk-corona models explain rather well various
Seyfert 1s iron line features already observed. However, these
models do not include the effect of spiral waves, which have been
already shown to be a natural outcome of the presence of magnetic
fields in the accretion disk. Note that magnetic fields must be
present to produce corona. Before closing, we emphasize that
inclusion of spiral waves in a realistic disk-corona model, as
shown in our current paper, gives rise to new features, such as
multiple fine peak structures, which can be tested by future
observations.

\acknowledgments

We are grateful to Darrell Rilett and Hideyo Kunieda for their
constructive comments on the manuscript.


\clearpage

\clearpage

\clearpage

\clearpage

\clearpage

\clearpage

\clearpage

\clearpage

\clearpage

\clearpage

\clearpage

\begin{deluxetable}{crrrrr}
\tabletypesize{\scriptsize} \tablecaption{Adopted Model
Parameters. \label{model}} \tablewidth{0pt} \tablehead{
\colhead{Model}& \colhead{$k_r$}  & \colhead{$\Delta_{sw}$} &
\colhead{$A_r$} & \colhead{Description of Model} &
\colhead{Figure} } \startdata
1 & 0.4 & 30 & $0.1$  & Mildly Packed & \ref{fig:profile-1-1}-(a) \\
2 & 1.0 & 30 & $0.1$  & Moderately Packed & \ref{fig:profile-1-1}-(b) \\
3 & 1.5 & 30 & $0.1$  & Tightly Packed  & \ref{fig:profile-1-1}-(c) \\
4 & 0.4 & 30 & $0.01$  & Rotation-Dominated\tablenotemark{a} &
\ref{fig:profile-1-1}-(d)
\\ \hline
5 & 0.4 & 5 & $0.1$  & Centrally-Concentrated 1 & \ref{fig:profile-2}-(a) \\
6 & 0.4 & 15 & $0.1$  & Centrally-Concentrated 2 & \ref{fig:profile-2}-(b) \\
1 & 0.4 & 30 & $0.1$  & Centrally-Concentrated 3 & \ref{fig:profile-2}-(c) \\
7 & 0.4 & 5 & $0.01$  & Rotation-Dominated\tablenotemark{a} &
\ref{fig:profile-2}-(d)
\\ \hline
1 & 0.4 & 30 & $0.1$  & Small Amplitude & \ref{fig:profile-3}-(a) \\
8 & 0.4 & 30 & $0.15$  & Intermediate Amplitude & \ref{fig:profile-3}-(b) \\
9 & 0.4 & 30 & $0.2$  & Large Amplitude & \ref{fig:profile-3}-(c) \\
10 & 0.4 & 30 & $0.02$  & Rotation-Dominated\tablenotemark{a} &
\ref{fig:profile-3}-(d)
\\ \hline

\enddata

\tablenotetext{a}{We set $A_r=0.1 A_{\phi}$ for these models.
Unless otherwise stated, $A_r=A_{\phi}$ and $\phi_{sw}=0\degr$ at
all the times.}
\end{deluxetable}

\end{document}